# A biomarker based on gene expression indicates plant water status in controlled and natural environments

Gwenaëlle Marchand [1,2], Baptiste Mayjonade [1,2], Didier Varès [1,2], Nicolas Blanchet [1,2], Marie-Claude Boniface [1,2], Pierre Maury [3,5], Fety Andrianasolo Nambinina [4,5], Philippe Burger [5,6], Philippe Debaeke [5,6], Pierre Casadebaig [5,6], Patrick Vincourt [1,2,7], Nicolas B. Langlade [1,2,7]

Corresponding author: Nicolas Langlade (e-mail : nicolas.langlade@toulouse.inra.fr)

**Addresses :**

1 INRA, Laboratoire des Interactions Plantes-Microorganismes (LIPM), UMR441, F-31326 Castanet-Tolosan, France

2 CNRS, Laboratoire des Interactions Plantes-Microorganismes (LIPM), UMR2594, F-31326 Castanet-Tolosan, France

3 Université Toulouse, INPT ENSAT, UMR1248 AGIR, F-31320 Castanet-Tolosan, France

4 CETIOM, Centre INRA de Toulouse, F-31320 Castanet-Tolosan, France

5 INRA, UMR1248 AGIR, F-31320 Castanet-Tolosan, France

6 Université Toulouse, INPT, UMR AGIR, F-31029 Toulouse, France

7 These authors contributed equally to this work.

## ABSTRACT

Plant or soil water status are required in many scientific fields to understand plant responses to drought. Because the transcriptomic response to abiotic conditions, such as water deficit, reflects plant water status, genomic tools could be used to develop a new type of molecular biomarker.

Using the sunflower (*Helianthus annuus L.*) as a model species to study the transcriptomic response to water deficit both in greenhouse and field conditions, we specifically identified three genes that showed an expression pattern highly correlated to plant water status as estimated by the pre-dawn leaf water potential, fraction of transpirable soil water, soil water content or fraction of total soil water in controlled conditions. We developed a generalized linear model to estimate these classical water status indicators from the expression levels of the three selected genes under controlled conditions. This estimation was independent of the four tested genotypes and the stage (pre- or post-flowering) of the plant. We further validated this gene expression biomarker under field conditions for four genotypes in three different trials, over a large range of water status, and we were able to correct their expression values for a large diurnal sampling period.

## KEYWORD INDEX

## INTRODUCTION

Water deficit in plants can be defined as the imbalance between the actual evaporative demand resulting from climatic conditions and the available water in the soil (Tardieu, Granier & Muller, 2011) This major environmental stress affects the growth and physiology of the entire plant and can therefore dramatically reduce crop yield and quality (Bhatnagar-Mathur, Vadez & Sharma, 2008). Recent intensification of drought events in Europe, Australia and North America, together with climatic model forecasts, suggest that drought will continue to dramatically affect crop productivity in the 21st century (Moriondo, Bindi, Kundzewicz, Szwed, Chorynski, Matczak, Radziejewski, McEvoy & Wreford, 2010). At the same time, the reduction of arable land area, the scarcity of water resources and the development of the human population all amplify the need to develop agrosystems that are more tolerant to water deficit or less water-consuming.

In this context, plant molecular physiologists, eco-physiologists and agronomists conduct experiments to estimate the plant response to water stress. One major requirement of these experiments is the estimation of soil water available to crops, at least at the most critical points of the developmental process. The direct estimation of water accessible to individual plants can be very difficult or even impossible in natural environments such as the field. Water status indicators can be of two types: soil- or plant-based measurements.

Soil-based measurements use either thermogravimetry, which requires samples for over-drying, or physical measurements of soil properties varying with its water content (Dobriyal, Qureshi, Badola & Hussain, 2012). Neutron probes have been widely used since the 1960s (reviewed by Gardner, Bell, Cooper, Dean, Gardner & Hodnett, 1991; Klenke & Flint, 1991), and dielectric methods, such as time-domain reflectometry (Topp & Davis, 1985) or capacitance sensors (Whalley, Dean & Izzard, 1992), have been used since the 1980s. Generally, these techniques can only describe limited soil regions that may not correctly represent the plant rhizosphere (Ferreira, Valancogne, Daudet, Ameglio, Pacheco & Michaelsen, 1996). The access tube used for the probe measurement is difficult to install and often modifies the water circulation and root dynamics surrounding the tube. Furthermore, these tools are time-consuming and labor-intensive and cannot be scaled up for high-throughput evaluation, which is needed in genetic analyses.

Therefore, plant-based measurements are often preferred. These measurements are based on the fact that the plant status reflects soil water availability. Morphological indicators can be used to evaluate drought stress. For example, breeders often score leaf rolling to estimate plant water status in monocots (O'Toole, Cruz & Singh, 1979). For perennial species, trunk diameter reflects water fluxes in the plant and can be used to manage irrigation (Goldhamer & Fereres, 2001). In eco-physiological studies, different indicators of plant status are commonly measured at different organizational levels, such as whole-plant transpiration, leaf water potential or stomatal conductance. For example, water transpiration cools the leaves and can therefore be monitored through thermal infrared measurements such as in wheat (Blum, Mayer & Gozlan, 1982). The measurement of pre-dawn leaf water potential ($\Psi_{PD}$) has been largely used for decades and is considered to be a standard. It is based on the fact that at dawn, water equilibrates between the rhizosphere and leaves, reflecting the water available to the plant. Therefore, it is subject to some limitations for heterogeneous soils (Ameglio, Archer, Cohen, Valancogne, Daudet, Dayau & Cruiziat, 1999) and its determination in numerous plants is restricted by the need of operating at pre-dawn.

Another method that is successfully used in controlled environments is based on the measurement of the daily transpiration of water-stressed plants relative to well-watered controls. This method defines a scale for water available for plant transpiration between an upper and lower limit of soil water content (SWC). The upper limit matches the SWC at field capacity and the lower limit is the SWC where relative transpiration decreases to less than 0.1 (Sinclair, 1986). According to Sinclair (2005), plants respond to the progressive drying of soil in a similar manner across a wide range of environmental conditions when this scale is used, where water stress is expressed as the fraction of transpirable soil water (FTSW). Variability in the leaf expansion and plant transpiration rate in response to water stress has been previously reported in different sunflower genotypes by using this method (Casadebaig, Debaeke & Lecoeur, 2008). When dealing with field conditions, several main drawbacks limit the applicability of this method. First the need for a control plot (in order to measure transpiration in well-watered conditions) doubles the experimental space required, which thus precludes the use of this indicator for large-scale genetic programs in crops. More importantly, to estimate FTSW in field environment, measurements of soil depth (soil profile, endoscopy), portion of soil explored by roots and soil water content (probes, soil cores) are required but are often unrealistic for genetic studies at microplot scale or poorly estimated.

In the context of the development of high-throughput phenotyping platforms, the need to develop a tool that would allow the early quantification of water deficit in a dose- and time-dependent manner is even more acute. Following the definition of Ernst & Peterson (1994) a soil water content biomarker could be a biochemical, physiological or morphological change in plants that measures their exposure to the environment (i.e., water deficit). This biomarker could therefore be used to reveal the status and trends in environmental assessment and also to predict crop responses to other biotic and abiotic stresses that interact with drought.

The transcriptomic response to water deficit is a widely described molecular process that allows plants to adapt to the water imbalance between supply and demand, and to develop a large range of morpho-physiological changes (Shinozaki & Yamaguchi-Shinozaki, 2007). The gene regulation cascade begins from the composite molecular perception of the environmental signal (the biophysical water imbalance) and moves via signal transduction down to the level of enzymes and structural proteins to produce biochemical compounds, fluxes and developmental adaptations. In accordance, some transcript expression levels are correlated only to FTSW but not to other major plant responses (Rengel, Arribat, Maury, Martin-Magniette, Hourlier, Laporte, Vares, Carrere, Grieu, Balzergue, Gouzy, Vincourt & Langlade, 2012). Such genes correspond to the definition of biomarkers that reflect soil water status.

Assembling several genes robustly correlated to soil water content appears to now be an attainable goal given recent progress in the description of the transcriptomic response to water deficit at the interface of molecular biology and eco-physiology (Ingram & Bartels, 1996); Ramanjulu & Bartels, 2002; Bartels & Sunkar, 2005; Harb, Krishnan, Ambavaram & Pereira, 2010; Aasamaa & Sober, 2011).

In fact, gene expression biomarker search and development is a long-standing goal of the reference meta-analysis platform Genevestigator (Zimmermann, Laule, Schmitz, Hruz, Bleuler & Gruissem, 2008). Recently, a successful meta-analysis of a large transcriptomic data set in maize allowed the development of a composite gene expression scoring system to quantitatively assess the response of maize to nitrogen conditions (Yang, Wu, Ziegler, Yang, Zayed, Rajani, Zhou, Basra, Schachtman, Peng, Armstrong, Caldo, Morrell, Lacy & Staub, 2011). Importantly, this first gene expression biomarker for *in planta* nitrogen status is independent of genotype, does not vary throughout plant development and was validated in field and greenhouse conditions.

The development of such tools has certainly been hampered by the rapid variation of plant transcriptome in response to many external factors, such as illumination and handling/wounding, as well as internal factors, such as the circadian clock. However, whole transcriptomic studies now show that part of the transcriptome robustly reacts to the plant environment in a dose- and time-dependent manner, which allows statistical models to be built, notably for the sunflower (Rengel *et al.*, 2012).

In this context, we used the sunflower as a model to develop a composite gene expression biomarker that is independent of genotype, developmental stage and time of day and that allows the estimation of soil water constraint in greenhouse and field experiments. This biomarker was standardized using the pre-dawn leaf water potential, FTSW, soil water content and fraction of total soil water when available.

# MATERIAL AND METHODS

## *Plant material and growing conditions*

Four experiments, i.e., one in greenhouse conditions and three in field conditions, using the four sunflower (*Helianthus annuus L.*) inbred lines XRQ, PSC8, their F1 named Inedi and another cultivated hybrid Melody were conducted in 2012 near Toulouse (Haute-Garonne, France).

For the greenhouse experiment conducted from May to June 2012, bleach-sterilized seeds were germinated on Petri dishes with Apron XL and Celeste solutions (Syngenta, Basel, Switzerland) for 3 days at 28°C. Plantlets were transplanted in 236 individual pots, and each pot contained one single plant.

Pots were filled with 15 L of a substrate composed of 10% sand, 40% P.A.M.2 potting soil (Proveen distributed by Soprimex, Chateaurenard, Bouches-du-Rhône, France) and 50% clay loam from the INRA site in Auzeville-Tolosane (Haute-Garonne, France).

Plantlets were sown on two different dates to obtain plants at two different stages (before and after flowering) respectively, 10 weeks and 4 weeks before the beginning of the stress treatment.

For each phenological (pre-flowering or post-flowering) stage with, respectively, 144 and 92 plants, the pots were arranged in a split-split-plot design with three blocks. The stress intensity (i.e., FTSW values of 0.8, 0.7, 0.5, 0.35, 0.20 and 0.12) was the main factor within the block, the genotype was the second factor, and finally the treatment (control plants were well-watered and treated plants were water-deprived) was the third factor. After an assessment of Alternaria blight evolution, plants of genotype PSC8 were not considered in the experiment dedicated to the post-flowering stage. Each pot was fertilized and irrigated as in Rengel *et al.* (2012) before the beginning of the water stress application.

In field conditions, the same genotypes were sown at three different locations: Samatan (Gers, France),

Fleurance (Gers, France) and Auzeville-Tolosane. In Samatan, the four genotypes were sown on April 20, 2012 and grown without irrigation. In Fleurance, PSC8, Melody and Inedi were sown on April 6, 2012 and grown without irrigation. In Auzeville-Tolosane, Inedi and Melody were sown on May 25, 2012 and grown in both irrigated (163 mm) and non-irrigated conditions.

The field experiments were designed in six randomized blocks for each location or location*condition combination. Each plot consisted of 12 rows with a length of 10 m, 12 rows with a length of 6 m and 9 rows with a length of 5.2 m for each genotype in Samatan, Fleurance and Auzeville-Tolosane, respectively, at the same plant population density (6.5 plants.m$^{-2}$).

Soil analysis

An 800-g soil sample was taken at depths of 60 cm and 30 cm in each trial and sent to the INRA LAS laboratory (Arras, Pas-de-Calais, France) for physical and chemical analyses.

Water stress treatment

In the greenhouse experiment, the pots were saturated with water 31 and 73 days after germination, respectively, for pre-flowering and post-flowering plants. The following morning, excessive water was drained for two hours and pots were weighed to obtain the saturation mass. From this point, irrigation was stopped for water-deprived (WD) plants. Both control and WD plants were weighed every day between 16:00 and 17:00 to determine the daily evapotranspiration. The water lost was added back to the control plants. To prevent soil evaporation, pots were covered with a 3-mm-thick polystyrene sheet. However, soil evaporation could not be neglected.

In the field experiments, 53, 70 and 40 mm of water were provided, respectively, on June 29, July 11 and August 13 2012 in the irrigated condition in Auzeville-Tolosane.

Soil evaporation estimation in the greenhouse

Six pots without plants that represented the different water content were also weighed every day for two weeks during the experiment. Climate conditions, such as relative humidity and average temperature in the greenhouse, were monitored daily. The soil evaporation could, therefore, be estimated by performing a linear regression with pot water content, relative humidity and average temperature using the function *regress* (MATLAB version 7.13.0.564, Statistics Toolbox 7.6).

This model is detailed in Supporting Table S1 and was used to estimate the soil evaporation during the greenhouse experiment.

Plant leaf area and transpiration in the greenhouse

For all plants, the length and width of odd leaves were measured every other day. The total leaf area was calculated from these measurements as described in Casadebaig *et al.* (2008)

The plant transpiration (E in $g.mm^{-2}$) for each pot was calculated every day as the difference between the water lost by the pot and the water lost by soil evaporation divided by the total plant leaf area.

The normalized transpiration (EN) for each WD plant was calculated every day as the ratio between its transpiration and the average transpiration of control plants of the same genotype in the same block.

## *Estimation of the fraction of transpirable soil water (FTSW) in the greenhouse experiment*

The total transpirable soil water (TTSW) is the maximum amount of soil water available to the plant. In our experiment, 8 treated plants reached EN values less than or equal to 10% and were used to estimate the TTSW. This weight ($W_{10\%}$) corresponded to the dry soil plus 26% (w/w) of the water contained in the saturated pot.

The fraction of transpirable soil water (FTSW) was finally calculated as follows:

FTSW = ($W_d$-$W_{10\%}$)/TTSW, where $W_d$ is the weight of the pot at day *d*.

The FTSW value was used to determine whether a plant had reached the target stress intensity.

## Estimation of the soil water content (SWC) in the greenhouse experiment

At the end of the greenhouse experiment, a soil core was collected from each pot. The soil samples were weighed to obtain the fresh weight and then dried for 48 h at 120°C before a second weighing to obtain the dry weight. The soil water content (SWC in percentage w/w) was calculated daily as follows: $SWC_j$ = ($W_{fi}$-$W_d$)/($W_d$-$W_d$'), where $W_{fi}$ is the weight of the fresh soil and plant at day *i*, $W_d$ is the weight of the dry soil and plant and $W_d$' is the weight of the dry plant. The weight of the fresh plant at day *i* is estimated using the dry weight of the plant and based on the assumption that the plant water content was, on average, 81% for post-flowering plants and 87% for pre-flowering plants and was the same for the different plant tissues.

## Estimation of the fraction of total soil water (FtotSW) in the greenhouse experiment

Another soil water content indicator is the fraction of total soil water (FtotSW), which was estimated as follows: FtotSW=($W_{fi}$-$W_d$)/($W_{sat}$-$W_d$), where $W_{sat}$ is the weight of the water-saturated pot.

## Measurement of leaf water potential (Ψ) in greenhouse and field experiments

In the greenhouse experiment, the harvested plants for transcriptomic analysis (WD and control plants) were placed in a dark room until the next morning. The water status at the time of harvest (between 11:00 and 12:30) was noted $\Psi_{PD}$' and estimated as the leaf water potential, after equilibrium with the soil was reached. $\Psi_{PD}$' was measured on the $n^{th}$ leaf for each harvested plant using a Scholander's pressure chamber (Soil Moisture Equipment Corp., California, U.S.A.), where n was 2/3 of the total leaf number $N_{tot}$.

In the field experiments, the water status at dawn was estimated as the classical pre-dawn leaf water potential $\Psi_{PD}$ and was measured for one plant per plot between 4:00 and 5:30 once a week for three weeks. Measurements began when plants were at the F1 stage (CETIOM nomenclature). In the Fleurance and Samatan trials, the first measurement occurred, respectively, on July 18 and 19, 2012. In Auzeville-Tolosane, the measurements began on July 31, 2012. The water potential was measured for the $5^{th}$ leaf from the head ($N_{tot}$ -5) using a Scholander's pressure chamber.

It is important to note that contrary to the transcriptomic harvests that were always performed at noon (except for the diurnal variation study), $\Psi_{PD}$ and $\Psi_{PD}$' are slightly different measurements of plant water status. $\Psi_{PD}$ was measured at dawn (between 4:00 and 5:30) after a normal night, thus representing soil-plant water status the day before leaf harvest (Fig S2 A). In contrast, $\Psi_{PD}$' was measured after the soil-plant water equilibrium had been reached during artificial night in dark room representing the exact plant water status at leaf harvest time (between 11:00 and 12:30) (Fig S2 B).

In the Auzeville-Tolosane trial, to study the influence of the diurnal variations on leaf water potential and gene expression, leaves were harvested in each of 3 blocks for each genotype, both in irrigated and non-irrigated conditions, in the same order, between 4:00 and 5:30, 7:00 and 8:30, 10:00 and 11:30, 11:30 and 13:00, 13:00 and 14:30, 16:00 and 17:30, 19:00 and 20:30, 22:00 and 23:30, and 1:00 and 2:30 (Fig S2 C). Separate leaves were used from the same plant for the leaf water potential measurement and for transcriptomic analysis. This study occurred on August 9 and 10, 2012 under high evaporative demand and constant sunny conditions.

## *Transcriptomic analysis*

<u>Selection of genes</u>

### *Gene indicators of water status*

From the study of Rengel *et al.* (2012), we selected genes that were found to (1) be correlated to the integrated transpired water (ITW) in fixed duration stress and in fixed intensity stress ($R^2 > 0.65$) and (2) show a treatment or a treatment*genotype interaction effect in the ANOVA for either type of stress. From this list of 143 genes, we chose to focus on sunflower homologues of Arabidopsis genes that have been described in the literature to be involved in abiotic stress responses. We finally kept 28 genes. Detailed descriptions of these genes are presented together with all the genes in this study in Supporting Table S3.

### *Circadian clock-related genes*

Numerous genes have been identified to vary according to the circadian clock. We chose the four Arabidopsis circadian clock regulators *TIMING OF CAB EXPRESSION 1 (TOC1)*, *LATE ELONGATED HYPOCOTYL* (*LHY*), *CONSTANS* (*CO*) and *ZEITLUPPE* (*ZTL*) (Alabadi, Oyama, Yanovsky, Harmon, Mas & Kay, 2001; Wilkins, Brautigam & Campbell, 2010) and identified the best BLAST hits in the sunflower transcriptome (https://www.heliagene.org/HaT13l). *HaDHN1* and *HaDHN2* of the sunflower were first described by Cellier, Conejero & Casse (2000) to vary during the circadian cycle and were re-examined in this study. All correspondences are summarized in Supporting Table S3.

### *Genes showing a genotype*treatment interaction effect in ANCOVA*

In addition to the 28 genes correlated to water stress intensity, we studied the gene expression levels of

four transcripts: HaT13l002164, HaT13l009999, HaT13l009995 and HaT13l020030. The expression of these transcripts was found to be correlated to three other morpho-physiological variables in Rengel *et al.* (2012): carbon isotopic discrimination (CID), evapotranspiration (ET) and osmotic potential (OP). The identification of Arabidopsis homologs was performed according to the best BLAST hits. These four genes and a fifth gene originally correlated to ITW in Rengel *et al.*, (2012) study, were used to illustrate a genotype*WSB interaction effect in ANCOVA analysis explained below. Detailed descriptions of the corresponding genes are presented in Supporting Table S3.

Primer design

Primers were designed using the HaT13l transcript sequence and Primer3 web tool (http://probes.pw.usda.gov/batchprimer3/index.html) using the default parameters with an optimal product size of 60bp (min=50bp, max=80bp). All primers are summarized in Supporting Table S4.

Tissue harvest and RNA extraction

One non-senescent and non-growing leaf by plant was harvested between 11:00 and 13:00, except during the diurnal variation study. Each leaf was sampled and treated separately. After freezing and grinding the samples, RNA was extracted and checked for quality and quantity. Detailed protocol of these steps and the cDNA synthesis is provided in Supporting Material and Method.

Estimation of gene expression by qRT-PCR

Gene expression was estimated by qRT-PCR. and was normalized according to the amplification efficiency and the expression levels of seven reference genes identified in Rengel *et al.* (2012). Detailed protocol of these steps is provided in Supporting Material and Method and reference gene information is summarized in Supporting Table S3.

<u>Gene expression correction following the time of the day</u>

The linear regression of gene expression as a function of the hour of the day was performed on the diurnal variation study's data between 10:00 and 20:00 for genes chosen for the biomarker model using the *robustfit* function in MATLAB. The linear regression was set to pass by means of the expression levels of samples harvested between 11:00 and 12:00, to match with the harvest time observed in the field and greenhouse experiments that were used to calibrate and validate the biomarker models. We corrected the gene expression for samples harvested at different times of the day to obtain an estimated gene expression at 11:30 using linear regression parameters.

## *Statistical analysis*

<u>Test of genotypic effect on the models</u>

For each selected gene correlated with water stress intensity, we performed a covariance analysis (*aoctool* function in MATLAB) by testing genotype-dependent (1) and non-genotype-dependent (2) models for the gene expression level as a function of water stress status as follows:

(1)   $Y_{i,t}=a_i+b_iX_{i,t}+Z_{it}$ : genotype-dependent model and

(2)   $Y_{i,t}=a+bX_{i,t}+Z'_{it}$ : genotype-independent model,

where $Y_{i,t}$ is the expression level of the gene for genotype *i* and for the actual water stress intensity *t* (with different values in each of the three blocks), $X_{i,t}$ is the value of the stress intensity, and $Z_{i,t}$ and $Z'_{i,t}$ are the residues.

The gene expression was considered to not have no genotypic effect if the F-test performed as follows was not significant (p>0.01):

$$F = (SSE_2 - SSE_1/(2*(G-1))/(SSE_1/df_1),$$

where $SSE_1$ and $SSE_2$ are, respectively, the sum of the squared errors for model (1) and model (2), G is the number of genotypes and $df_1$ is the number of degrees of freedom attached to the error in the model (1).

<u>Statistical calibration and validation of the water status biomarker (WSB)</u>

All combinations of 3, 4, 5 or 6 genes that were correlated to stress intensity and presented no genotypic effect were tested to construct a model to estimate the pre-dawn leaf water potential. The model fitting was performed by the *GeneralizedLinearModel.fit* function in MATLAB using the greenhouse data as the calibration set. For each of the four types of models, we selected the 50 best models according to the AIC criterion.

Selected models were then tested using the *predict* function in MATLAB, and field data served as the validation set. For each of the four types of models, we selected the best model according to the $R^2$ of the correlation between $WSB_{\Psi PD}$ predictions and the corrected values of observed $\Psi_{PD}$ using the *regress* function in MATLAB (corrections are described in Supporting Material and Methods and Fig S5). We compared the four types of models and chose the best one according to the $R^2$ of the correlation.

## *Genes with Genotype*WSB interaction*

<u>Test of trial effect</u>

We considered samples harvested only in non-irrigated conditions for the three field trials. For the four genes correlated to morpho-physiological traits and and one gene to ITW, we performed an ANOVA using the *anovan* function in MATLAB to test the genotypic, trial and genotypic*trial interaction effects.

<u>Test of the genotype*WSB interaction</u>

Genes found to have no trial effect were tested for the genotype*WSB interaction. We performed a covariance analysis using the *aoctool* function in MATLAB with the following model:

$$Y_{i,b} = a_i + b_i X_{i,b} + Z_{i,b},$$

where $Y_{i,b}$ is the expression level of the gene for genotype *i* and biomarker level *b*, $X_{i,b}$ is the value of the WSB level and $Z_{i,b}$ is the residue.

# RESULTS

## *Greenhouse results*

<u>Selection of candidate genes</u>

Based on our previous results (Rengel *et al.*, 2012), we selected 28 genes that were found to be correlated to the integrated transpired water (ITW) in fixed duration stress and in fixed intensity stress ($R^2>0.65$). As the expression of these genes was independent of the tested genotypes, they were strong candidates to build a biomarker for plant water status. To assess a particular level of gene expression that reflects stress intensity, we needed to study these genes through a larger range and at a finer scale of drought stress.

<u>Establishment of a fine scale of drought stress</u>

To study changes in gene expression at different stress levels, we established a large range of drought stress with a fine scale. For the treated plants, the water status indicators ranged from 0.97 to -0.087 for the FTSW, from -0.2 to -2.4MPa for the pre-dawn leaf water potential ($\Psi_{PD}$'), from 54.3% to 5.98% for the soil water content (SWC) and from 0.13 to 1 for the FtotSW. The four genotypes and the two stages were represented through the entire range.

The four water status indicators measured during the greenhouse experiment were highly correlated with the $R^2$ values, ranging from 0.65 to 0.96 (Fig. 1). Interestingly, the $\Psi_{PD}$' was only correlated with FTSW values below 0.4, SWC values below 25% and FtotSW values below 0.5. This selective correlation reflects that, in our data, $\Psi_{PD}$' did not discriminate high water status levels.

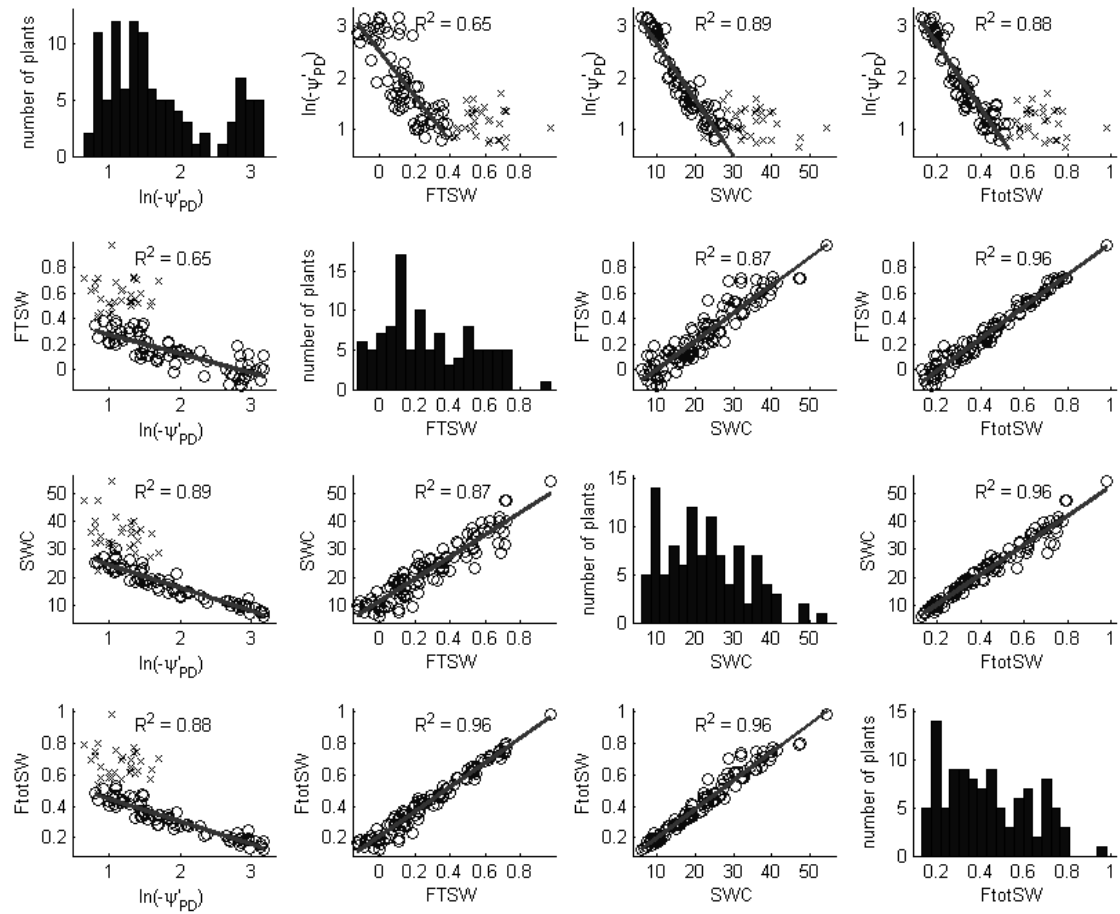

<u>Figure 1</u>: The distributions and correlations between the four water status indicators measured in the greenhouse experiment: ΨPD', FTSW, SWC and FtotSW.

<u>Correlation between gene expression and water status indicators</u>

As a confirmation of our previous results using the FTSW (Rengel *et al.*, 2012), we estimated the correlations of 28 selected genes to the four water status indicators over the new finer and larger scales of water status, considering together the treated plants of all genotypes and growth stages. Raw data of the gene expression level compared to FTSW level for the 28 candidate genes are shown in figure S6. We identified 18 genes whose expression was correlated (p<0.01) to the FTSW, 20 for $\Psi_{PD}$', 21 for the SWC and 18 for FtotSW (Table 1 and Fig S7).

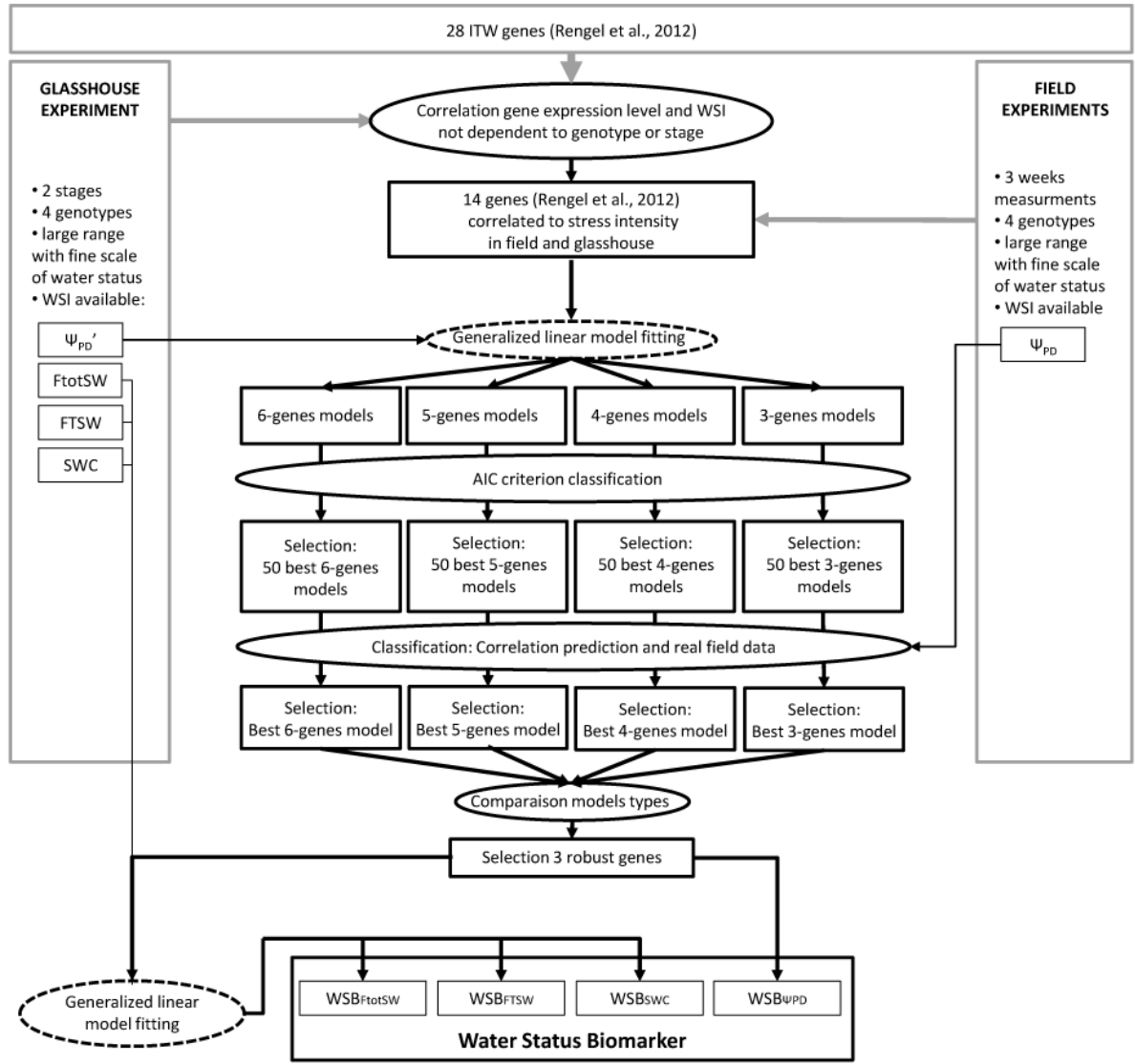

Figure 2. A schematic description of the water status biomarker construction. WSB$_{\Psi PD}$ was developed in greenhouse conditions and validated in the field. WSB$_{FtotSW}$, WSB$_{FTSW}$ and WSB$_{SWC}$ were built in the greenhouse environment using field-robust genes used for WSB$_{\Psi PD}$.

A covariance analysis was used to test genotype-dependent correlations (p<0.01). Among the correlated genes, we found two genes (according to the water status indicator) whose correlations were genotype-dependent (summarized in Table 1 and Fig S8).

Finally, among the 28 initial genes, we retained 14 genes that showed neither genotype nor stage effects in the greenhouse experiment and that were technically robust in both greenhouse and field experiments. These first steps of gene selection for biomarker construction are summarized in Figure 2.

|  | FTSW | $\Psi_{PD}$' | SWC | FtotSW |
|---|---|---|---|---|
| Number of genes correlated to water indicators | 18 $0.22<R^2<0.75$ | 20 $0.19<R^2<0.91$ | 21 $0.18<R^2<0.0.8$ | 18 $0.18<R^2<0.77$ |
| Number of genes correlated to water indicators without genotype effect | 16 $0.22<R^2<0.71$ | 19 $0.19<R^2<0.83$ | 21 $0.18<R^2<0.77$ | 21 $0.18<R^2<0.8$ |

**Table 1**. The number of genes correlated (p<0.01) to each water deprivation indicator and genotype effects to be used in gene combinations for biomarker fitting

## Construction of Generalized Linear Models to estimate plant water status

<u>Model for $\Psi_{PD}$' in glasshouse</u>

According to the AIC criterion, we selected the 50 best generalized linear models with 3, 4, 5 and 6 genes to fit the $\Psi_{PD}$' from the greenhouse data (Fig 2). The adjusted $R^2$ and RMSEc for the different types of models are presented in Table 2. Considering only these glasshouse data, the adjusted $R^2$ increased with the number of genes introduced in the model, and the RMSEc values for the four types of models were similar.

|  | 3 gene models | 4 gene models | 5 gene models | 6 gene models |
|---|---|---|---|---|
| Adjusted $R^2$ | 0.73-0.82 | 0.80-0.83 | 0.82-0.85 | 0.84-0.86 |
| RMSEc | 0.61-0.66 | 0.64-0.66 | 0.64-0.67 | 0.65-0.68 |

**Table 2**. The range of adjusted $R^2$ and RMSEc for the 50 best linear models with 3, 4, 5 and 6 genes fitting the $\Psi_{PD}$' in the greenhouse experiment.

Field experiment validation with $\Psi_{PD}$

We used the results of three field trials to select the best predictive model for $\Psi_{PD}$' (Fig 2). The field trial experiments were implemented in environments with deep (Auzeville-Tolosane) or shallow (Fleurance and Samatan) soils. The Fleurance and Samatan trials had clay soils (respectively in average 52.5% and 52.8% of clay) with a low water-holding capacity. The Auzeville-Tolosane trial had soil with an equilibrate texture between the silt loam and sand (composed in average of 24.3% of clay, 29.8% of silt and 45.9% of sand), and therefore, with a high water holding capacity and therefore with a high field capacity (Supporting Table S9). Trials were chosen with different soil characteristics to ensure a wide range of plant water statuses. For the same reason, we harvested samples and measured $\Psi_{PD}$ over 3 weeks for six repetitions per genotype. Finally, the Samatan data over the last week was disturbed by an important rain event and was discarded. Overall, we obtained a good range of water stress across the entire experiment: $\Psi_{PD}$ ranged from -0.5 to -2.2 MPa in the Fleurance trial and from -0.7 to -2.3 MPa in the Samatan trial, whereas in Auzeville-Tolosane, where we set up irrigated and non-irrigated conditions, $\Psi_{PD}$ ranged from -0.3 to -1.5 MPa.

Using field experiment data, we compared the models' prediction (WSB$_{\Psi PD}$) and $\Psi_{PD}$' estimated from the measured $\Psi_{PD}$. We observed that models with 6 genes that were better in the greenhouse environment introduced errors in field predictions. We selected the three-gene model that showed the best correlation between the observed and predicted $\Psi_{PD}$' with an $R^2$ equal to 0.61 (Fig 3) and an RMSEp of 0.67. With this model, the WSB$_{\Psi PD}$ was estimated as follows:

WSB$_{\Psi PD}$=ln(-$\Psi_{PD}$') = 1.53 + 0.35*ln(dCt$_{HaT131002207}$) − 0.39*ln(dCt$_{HaT131002636}$) +0.16*ln(dCt$_{HaT131s5199}$).

where $\psi_{PD}$' is expressed in 0.1MPa.

In the greenhouse, this model had an adjusted $R^2$ of 0.78 and an RMSEc of 0.64 and therefore offered better prediction in both controlled and field environments.

Models for water stress indicators not accessible in field conditions

The three genes used in the model to predict $\Psi_{PD}$' appeared to be robust enough in predicting the stress intensity in both the greenhouse and field environments. We used these same genes in the construction of models for water stress indicators that are not accessible in field conditions. FTSW, SWC and FtotSW were estimated by generalized linear modeling using gene expression levels of HaT13l002722, HaT13l002636 and HaT13l005199 as follows:

$$WSB_{FTSW} = 0.42 - 0.0618*\ln(dCt_{HaT13l2207}) + 0.21*\ln(dCt_{HaT13l002636}) - 0.04*\ln(dCt_{HaT13l005199}),$$

$$WSB_{SWC} = 27.70 - 3.83*\ln(dCt_{HaT13l002207}) + 8.51*\ln(dCt_{HaT13l002636}) - 1.79*\ln(dCt_{HaT13l005199}), \text{ and}$$

$$WSB_{FtotSW} = 0.54 - 0.06*\ln(dCt_{HaT13l002207}) + 0.17*\ln(dCt_{HaT13l002636}) - 0.03*\ln d(Ct_{HaT13l005199}).$$

Models had adjusted $R^2$ values of, respectively, 0.69, 0.72 and 0.74 for FTSW, SWC and FtotSW, and their RMSEc values were, respectively, 0.20, 9.31 and 0.18.

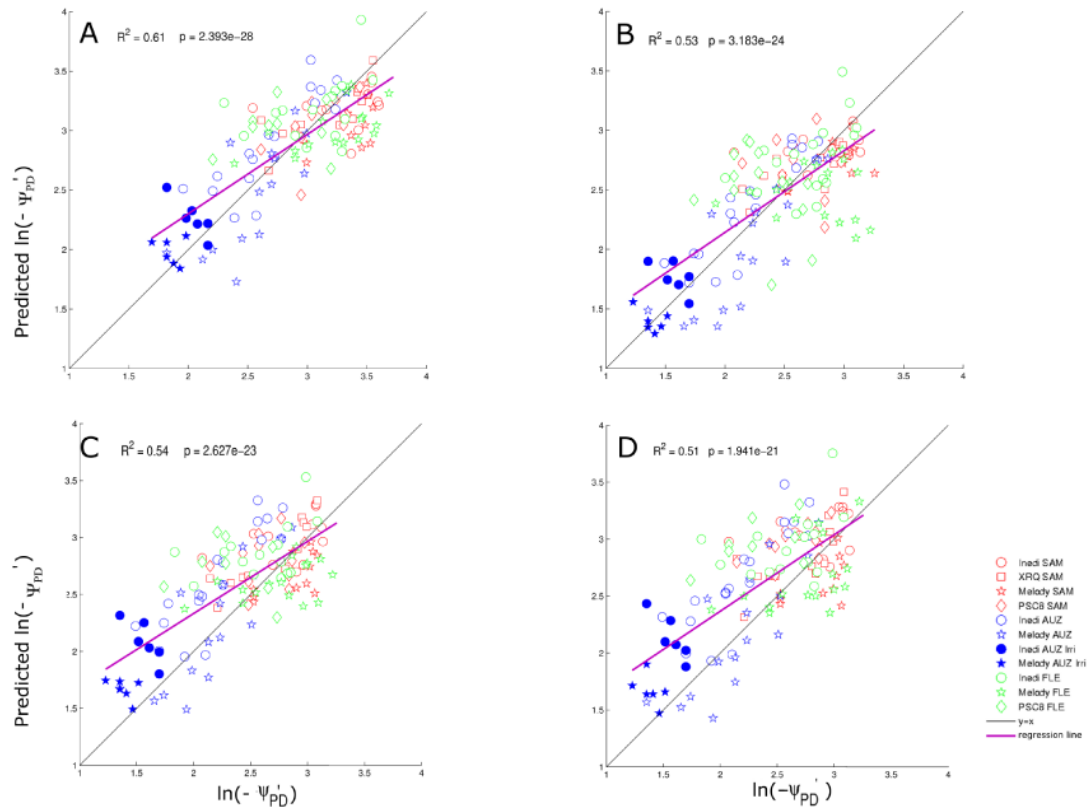

<u>Figure 3</u>. Correlations between corrected field data $\ln(\Psi_{PD}')$ and prediction ($WSB_{\Psi PD}$) of the corresponding best model with 3, 4, 5 or 6 genes. A, Correlation with predictions of the best three-gene model. B, Correlation with predictions of the best four-gene model. C, Correlation with predictions of the best five-gene model. D, Correlation with predictions of the best six-gene model. Field data of the three trials are represented: Samatan (SAM, in red), Fleurance (FLE, in green) and Auzeville-Tolosane (AUZ, in blue). Note that the three-gene model, represented by the regression line in violet, produced better predictions, with an $R^2$ value of 0.61.

## Correction of gene expression level for diurnal variation

To build these models, we used samples harvested between 11:00 and 12:30. So, the biomarker was calibrated and validated only for samples harvested during this period of the day. To use this model as a biomarker and a practical tool in experiments involving large numbers of genotypes or conditions, it appeared to be useful to obtain a biomarker valid for a larger sampling time period. Therefore we needed to correct for the time of the sampling, at least for genes showing modification of their expression according to the diurnal variation. This variation of expression of the three genes included in the models (shown in Fig 4A-C) throughout a 24-hour period could not be neglected in comparison to the variation of known circadian genes (Supporting Figure S10). The kinetic curves of gene expression levels over 24 hours showed that between 10:00 and 20:00, the variation of gene expression could be estimated through a linear regression. We used kinetic curves over 24 hours to estimate the expression at 11:30 from the expression at any time over this specific timeframe as shown in Figure 4DE. The correction was efficient for samples harvested from 10:00 to 17:30; however, for samples harvested out of this timeframe, the correction was not sufficiently reliable to estimate the gene expression at 11:30. Sampling out of this timeframe should therefore be avoided.

| Gene | Genotype effect (p-value) | WSB effect (p-value) | Genotype*WSB interaction effect (p-value) |
|---|---|---|---|
| HaT13l009400 | 4.98E-12 | 1.62E-17 | 1.90E-02 |
| HaT13l002164 | 9.00E-04 | 1.52E-04 | 3.90E-04 |
| HaT13l009999 | 8.35E-05 | 1.25E-08 | 1.75E-02 |
| HaT13l009995 | 5.70E-01 | 6.37E-08 | 4.83E-0.2 |
| HaT13l020030 | 3.41E-05 | 3.78E-03 | 7.23E-03 |

**Table 3**. Results of covariance analysis for five selected genes. These genes shown G*WSB interaction effect (p<0.05) and illustrate the use of the biomarker to detect differential plant drought responses according to the genotype and the plant water status as it is identified by the WSB.

## *Use of the Water Status Biomarker*

Identification of gene expression profiles showing a genotype*WSB interaction in field conditions

The water status biomarker (WSB) built in this study could be applied to characterize the environments for water stress for different genotypes. This application would allow the identification of genes showing genotype*water status interactions that could explain the genotypic variation for drought tolerance. To illustrate this, we chose five genes correlated to other morpho-physiological variables or water stress intensity in Rengel *et al.* (2012), that did not show a trial effect in field experiments (p>0.05 in ANOVA over the three non-irrigated trials) as summarized in Table S11. For these five genes, a covariance analysis showed significant genotypic and genotype*WSB interaction effects as shown in Table 3 and Figure 5. These results exemplify a possible use of the WSB when searching for genetic variation of the drought response.

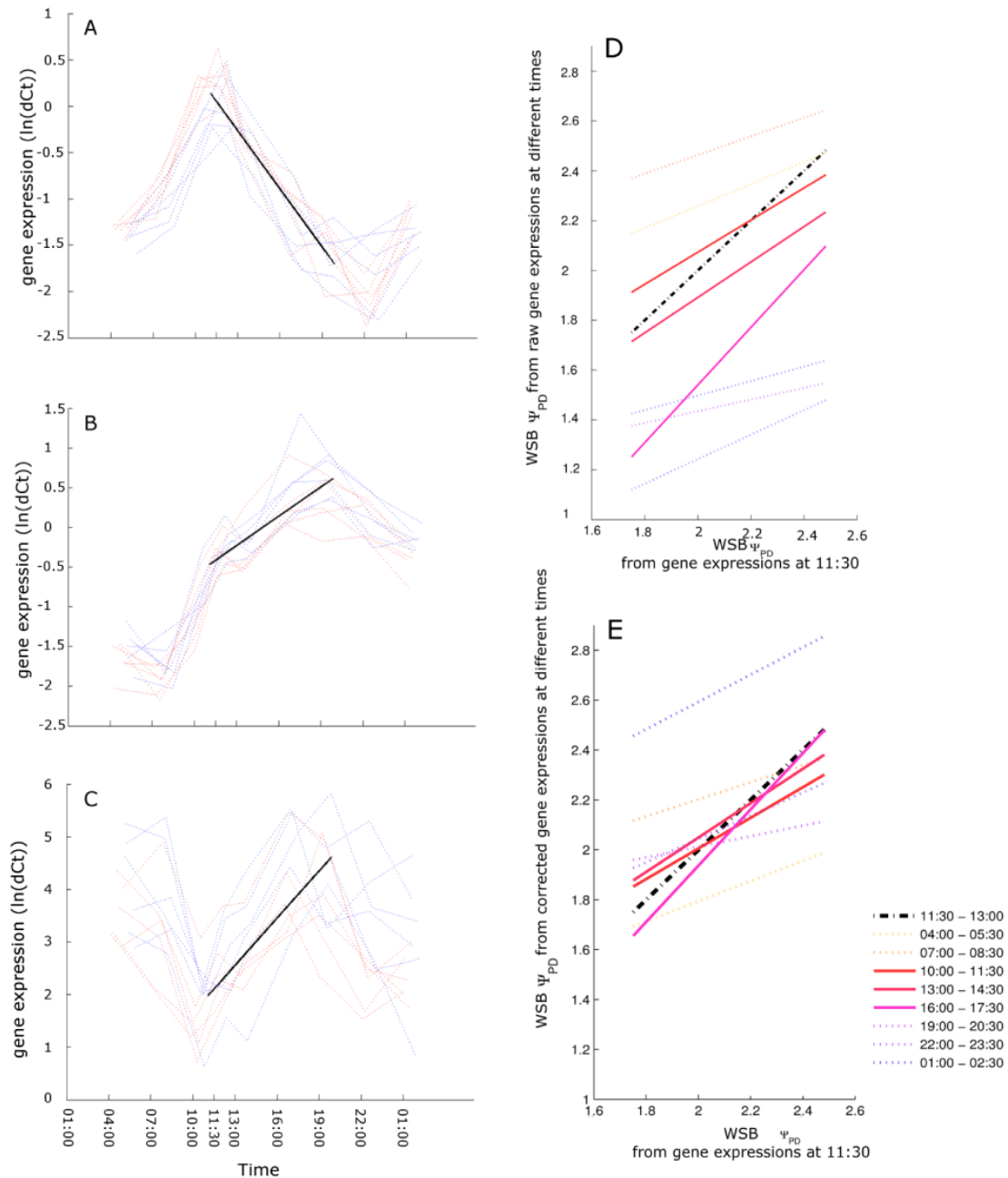

Figure 4.   Twenty-four hour kinetic curves of expression levels for the three transcripts used in biomarker construction. And correction efficiency for biomarker prediction from samples harvested at different times. A, Kinetic curve for transcript HaT13l002207. B, Kinetic curve for transcript HaT13l002636. C, Kinetic curve for transcript HaT13l005199. Dotted lines represent kinetic curves for the 12 plots. The solid line is the regression line between 10:00 and 20:00 used for transcript expression correction. D, Comparison between biomarker predictions using gene expression at 11:30 and biomarker predictions using raw gene expression at different times. E, Comparison between biomarker predictions using gene expression at 11:30 and corrected gene expression at different times. Correction aimed at estimating gene expression at 11:30 from samples harvested between 10:00 and 20:00 and showing a linear variation. Note that the correction is efficient only for samples harvested between 10:00 and 17:30

# DISCUSSION

## *Description of the three genes selected for the water status biomarker*

The water status biomarker was defined from the expression levels of three genes normalized by the expression levels of reference genes. HaT13l002207 is homologous to the Arabidopsis transcript of *TUA5* (AT5G19780). This gene encodes a tubulin. Microtubules are polymers of tubulin heterodimers. The relationship between microtubules and ABA in plant cells has been extensively studied, although the exact mechanisms involving the microtubule response to drought stress remain largely unknown. Dynamic microtubules in guard cells are sensitive to extracellular stimuli and drought stress, which affect both the microtubule dynamics and ABA accumulation (Marcus, Moore & Cyr, 2001). Moreover, Lu, Gong, Wang, Zhang & Liang, (2007) demonstrated that changes in microtubule dynamics have an effect on ABA accumulation in root cells of *Zea mays*.

HaT13l005199 is homologous to the Arabidopsis transcript of *XTR7* (AT4G14130). This second gene encodes for a concanavalin that is a xyloglucan endotransglycosylase (XET). XETs form a large family, of which some members are involved in cell wall biogenesis or rearrangement (Van Sandt, Suslov, Verbelen & Vissenberg, 2007), which are processes inherent to growth. Moreover, a relationship between the response of the growth rate under water stress and XET activity has been previously suggested (Thompson, Wilkinson, Bacon & William, 1997).

HaT13l002636 is homologous to the Arabidopsis transcript of *GBF3* (AT2G46270). This last gene is a transcription factor that encodes a bZIP G-box binding protein, and its expression was found to be induced by ABA, cold and water deprivation (Lu, Paul, McCarty & Ferl, 1996).

These three genes were shown to have direct or indirect links with water deficit or ABA, which is the drought stress hormone. Although these links were demonstrated in Arabidopsis and in the microtubule dynamics of maize, our biomarker gene selection and model calibration may be specific to the sunflower.

Accordingly, new WSB development would be required for other species.

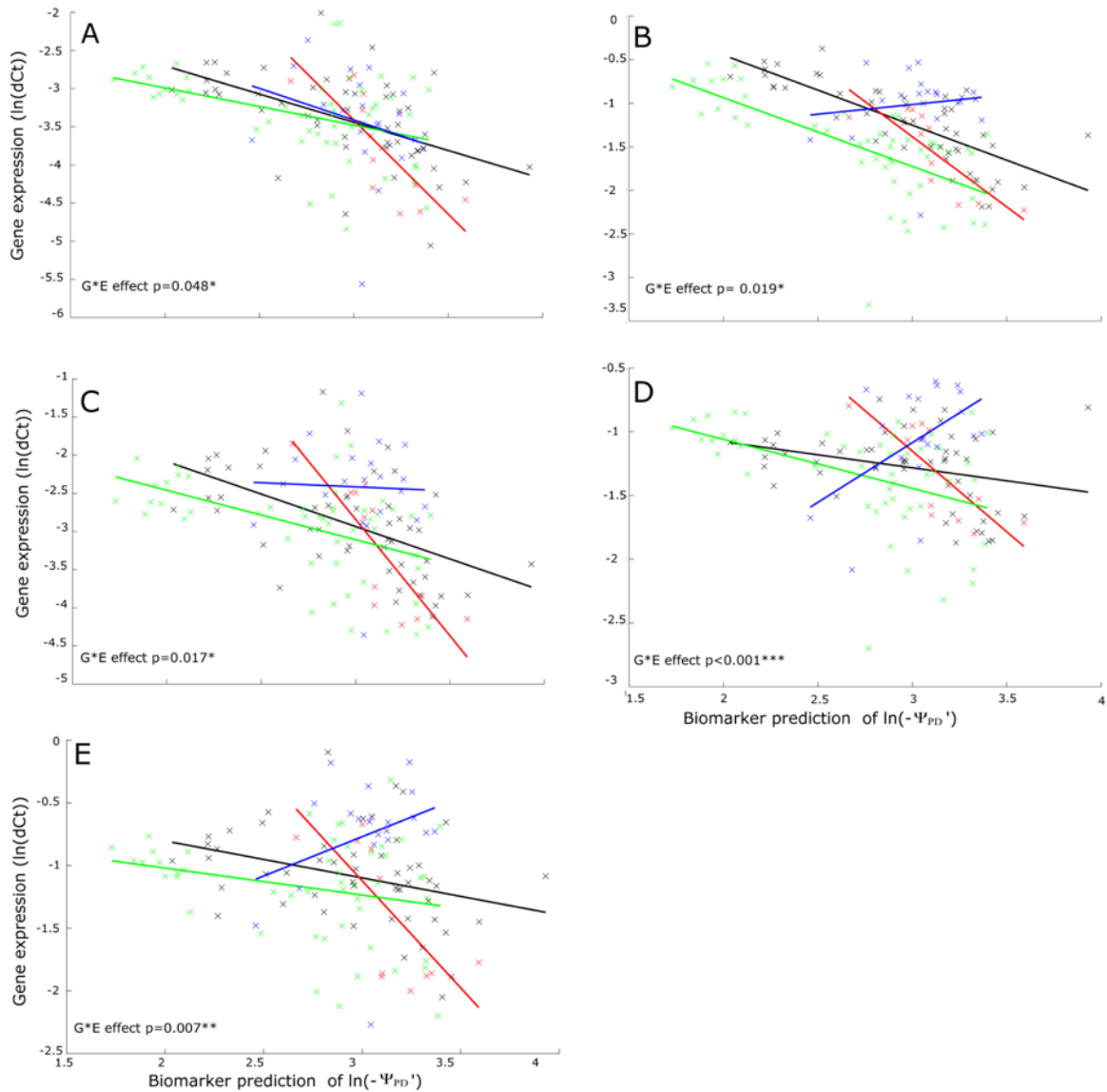

Figure 5. Gene expression showing genotype*WSB interactions. The four genotypes Melody (green), Inedi (black), PSC8 (blue), and XRQ (red) showed different interactions with the environment described by the WSB$_{\Psi PD}$. A, HaT13l009995, homologous to the Arabidopsis *TIL1* transcript. B, HaT13l004900, homologous to the Arabidopsis *UBC28* transcript. C, HaT13l009999, homologous to the Arabidopsis *SP1L4* transcript. D, HaT13l002164, homologous to the Arabidopsis *HAP2A* transcript. E, HaT13l020030, homologous to the Arabidopsis *AIG2L* transcript.

### *Comparison between the WSB and classical water status indicators*

From the correlations observed between the $\Psi_{PD}$', FTSW, SWC and FtotSW, we confirmed that plant-based water status indicators, such as $\Psi_{PD}$' and FTSW, reflect the soil water content. The expression levels of genes selected to construct the WSB were highly correlated to the water status indicators, especially to the plant-based indicator $\Psi_{PD}$'. Therefore, the expression of these three genes reflects the environmental water status as integrated by the plant. As the expression of these genes was independent of genotype and stage, the determination of a given drought stress based on the water status biomarker was the same for all genotypes and stages tested.

We obtained a better WSB for $\Psi_{PD}$ because we selected genes using the $\Psi_{PD}$' measurements but also because gene expression levels and $\Psi_{PD}$' are both plant-based measurements. Therefore, to characterize the water available for the plant in the field, WSB$_{\Psi PD}$ was the most robust biomarker.

### *Advantage of WSB over environmental data*

The WSB built in this study represents the environmental water status perceived by the plant. The expression of the selected genes was correlated to the environmental water status and independent of genotypic diversity and stage. However, the water status described by the biomarker was not exactly the same as the water status described by soil-based water indicators. For example, the SWC reflected the water status of the soil, but according to the type of soil, the availability of water to the plant could vary. As a plant-based water status indicator, gene expression levels have the advantage of offering a better representation of the environment perceived by the plant than raw soil and climatic data.

### *WSB genes and environmental signal integration*

Gene regulatory networks integrate the environment to drive morphological and physiological responses

(Shinozaki & Yamaguchi-Shinozaki, 2007). Any genotypic variability in a step of the response cascade would translate in further genotypic variations in the overall drought response process. Therefore, we can argue that the genes used in our genotype-independent biomarker are involved in upper steps of this cascade close to the integration of the environmental signal or in general genotype-independent pathways (maybe illustrated by the presence of the tubulin gene in our WSB).

## *Validity of the WSB*

The water status biomarker was developed using the correlation with $\Psi_{PD}$' in a greenhouse and validated with field $\Psi_{PD}$ data (Fig. 2). Because they are not easily tractable in field conditions, the estimation of FTSW, SWC and FtotSW using our gene expression biomarker was only validated in the greenhouse experiment. However, to account for greenhouse and field variations, we built $WSB_{FTSW}$, $WSB_{SWC}$ and $WSB_{FtotSW}$ with the same genes used for the $WSB_{\Psi PD}$, as they were shown to be robust in both environments. This robustness allows us to be confident in our predictions of the $WSB_{FTSW}$, $WSB_{SWC}$ and $WSB_{FtotSW}$ in the field and makes these indicators accessible in this environment; providing a new tool for plant biologists.

The circadian regulation of gene expression has been documented for a couple of sunflower genes (Cellier *et al.*, 2000). However, to construct the biomarker, we did not take into account that the diurnal variation of our genes expression could be important. To calibrate and validate the water status biomarker, we sampled plant tissues between 11:00 and 12:30, so the problem of sampling period was not crucial. With the diurnal variation study, we were also able to infer the gene expression level at 11:30 from samples harvested between 10:00 and 17:30, which thus defined a valid timeframe for sampling. This correction was specific to the day of study (sunny and warm) and might at least improve the prediction in most of the drought studies.

The biomarker construction was based on samples harvested from May 31 to June 15 2012, including average relative humidity varying from 63 to 80% on cloudy and sunny days reflecting variable evaporative demands. It selected genes that were not affected by this kind of climatic variations. This was confirmed in the field experiments performed two months later in shorter day conditions, in three different locations with higher evaporative demands.

In these conditions, the biomarker gene expression was not correlated to diurnal water potential and was not influenced by the specific climatic conditions of the day of harvest. So in all these different relative humidity and average temperature conditions, we could validate our model for both experimental conditions.

Both in the greenhouse experiment and in field trials, we managed to obtain a large range of water stress, and we showed that the WSB was able to characterize the environment for this entire range of water statuses. However, in all experiments, we only applied a continuous deficit of water. We did not test the ability of our WSB to describe plant recovery if, for example, rain events occur after a long period of water deficit.

Importantly, the expression of the three genes chosen to the build biomarker appears to be independent of the tested genotypic diversity. This independence is a particular characteristic of the selected genes and makes them distinct from many other genes as we describe below.

## *Use of the WSB*

Breeding for a trait affected by drought (G*E)

For crop breeding, environmental characterization is critical to understand the genotype*environment interaction. Climate and crop management data alone are not sufficient to obtain a good definition of the environment perceived by the plant. Therefore, the WSB developed in this study could be useful for characterizing the environment with regard to water availability, allowing breeders to better understand genotype*drought stress interactions. Accordingly, this WSB could be a powerful tool to study any trait affected by drought and help to breed drought tolerance in sunflower.

*Example: identification of gene expression responses depending on the water status biomarker*

Following this G*E identification strategy, we looked for genes showing significant genotype*WSB interaction effect and whose expression levels were independent of trials.

Our results suggest that five genes could show this pattern (Fig. 5). Among these genes, four were found correlated with morpho-physiological variables and the last one to water stress intensity in Rengel *et al.* (2012). These genes were examples of genes related to plant drought responses and whose expression changed according to the genotype and the plant water status as predicted by the WSB. The expression of HaT13l002164 and HaT13l009999 was correlated with carbon isotopic discrimination (CID). These transcripts are respectively homologous to the Arabidopsis transcripts of *HAP2A* (AT5G12840), which codes for a subunit of the CCAAT-binding complex, and *SP1L4* (AT5G15600), which regulates cortical microtubule organization. The third gene encodes HaT13l009995, whose expression is correlated with evapotranspiration (ET). This transcript is homologous to the Arabidopsis transcript of *ATTIL* (AT5G58070) involved in thermotolerance (Chi, Fung, Liu, Hsu & Charng, 2009). The fourth gene encodes HaT13l020030, whose expression is correlated with the osmotic potential (OP). Its Arabidopsis homolog is the avirulence-induced gene *AIG2L* (AT5G39720). Finally, HaT13l009400 expression was found to be correlated with water stress intensity but was not used to build the biomarker because it showed a genotype effect in the greenhouse experiment. This transcript is homologous to the Arabidopsis transcript of *UBC28* (AT1G64230), which codes for a ubiquitin-conjugating enzyme.

Following the signaling cascade from the environmental signals down to the genotype-specific responses,

these genes would be responsible for the final responses and belong to the end of the gene signaling cascade. Because we were able to identify environment-related genes and response-related genes, this approach could allow us to model the gene regulatory network from the global gene expression dataset.

Crop model

Crop models represent dynamic crop processes and are used to simulate crop development and behavior as a function of the environment, management conditions and genetic variations (Sinclair & Seligman, 2000). Such tools may also benefit from the use of the WSB. For example, SUNFLO (Casadebaig, Guilioni, Lecoeur, Christophe, Champolivier & Debaeke, 2011) is a crop model that is able to simulate biomass yield and transpiration of the sunflower genotypes in contrasting environments. In this model, the FTSW is an output variable of a water budget module based on climatic and management input variables and plant parameters (expansion and transpiration sensitivity to water stress, soil depth and water holding capacity). The simulated FTSW is thereafter used to model the effects of water stress on crop growth and performance.

In this context, $WSB_{FTSW}$ could be a tool to readjust the simulated FTSW values with observations to improve crop performance prediction for a specific site. It appears impossible to harvest plants every day to obtain a daily $WSB_{FTSW}$. However, harvesting at a few key stages of crop development appears to be a good compromise and could help to perform a more accurate simulation of crop development.

Distinction between traits of interest and drought stress responses

In the field, crops are actually subjected to both abiotic and biotic stresses. These two types of stresses are in interactions. A biomarker characterizing water status could be a tool to distinguish the part of genetic variation in a trait of interest, such as distinguishing the resistance to a disease from drought stress responses that could interfere in phenotyping. As an example, it has been shown that water deficit conditions were significantly involved in the disease severity of premature ripening induced by *Phoma macdonaldii* susceptibility (Seassau, Dechamp-Guillaume, Mestries & Debaeke, 2010). In this case, the biomarker for characterizing the water status environment perceived by the plant could be used to perform a screening of *Phoma*-tolerant genotypes adjusting for different water statuses.

Drought stress management feasibility: time and cost of the WSB

Our goal in this study was not only to demonstrate the possibility to characterize the water status environment from gene expression levels but also to design a practical tool that could be easily used. To achieve this goal, we paid particular attention to the cost and time needed to run the new biomarker.

Important parameters to consider in the development of a cost- and time-effective biomarker are the sampling time window and the number of genes used.

Concerning the time window for sampling, the diurnal variation study allowed us to estimate gene expression levels at 11:30 from samples harvested between 10:00 and 17:30. Therefore, the WSB can be used with samples harvested during a large diurnal sampling period, in contrast to the $\Psi_{PD}$, that can only be measured at pre-dawn.

Regarding the number of genes, we developed a WSB based on the expression of only a few genes, i.e., the three genes included in the generalized linear models and reference genes used for normalization. Therefore, it is possible to easily test very large numbers of samples using q-PCR with minimal time and cost.

However, because of the delay between harvest and q-PCR results, the WSB seems more relevant for breeding or studying genotype behavior than for drought stress management during crop production.

# CONCLUSIONS

In this study, we developed a gene expression biomarker that was able to estimate the plant water status expressed as the $\Psi_{PD}$', FTSW, FtotSW or SWC (Fig. 7). This tool is independent of the tested genotypes and the developmental stage. A correlation between the WSB and $\Psi_{PD}$' was validated in greenhouse and in field conditions with different soil properties. Other classical water status indicators showed robust correlations with the WSB in greenhouse experiments. The water status biomarker developed here could be a useful tool in different scientific fields for characterizing the water status in plants.

# ACKNOWLEDGEMENTS

This work was part of the OLEOSOL project funded by French public funds for competitiveness clusters (FUI), the European Regional Development Fund (ERDF), the Government of the Région Midi-Pyrénées, the Departmental Board of Aveyron (France), and the Cities Cluster of Rodez (France). It benefited from part of a PhD grant co-funded by Syngenta Seeds and the Région Midi-Pyrénées.

We deeply thank Loïck Aymard for his help in the greenhouse, field and laboratory parts of the project and Julien Berceron, Valentin Boniface, Kenneth Gamas and Richard Bonnefoy for soil sampling in the field. We thank the Auzeville-Tolosane INRA Experimental Unit, Soltis and Syngenta Seed companies for the field trial implementations. We thank the Platform GeT PlaGe  from INRA Toulouse for their technical support. We finally thank all the sunflower team members and the common services of the LIPM.

# AUTHOR CONTRIBUTIONS

G.M., D.V., P.M., P.D., P.V. and N.B.L. designed the research, G.M., B.M., D.V., B.N., M.-C.B., P.M.,

F.A.N., P.C. and N.B.L. performed the research, G.M., B.M., P.M., P.V. and N.B.L. analyzed data and

G.M., B.M., P.B., P.C., P.V. and N.B.L. wrote the manuscript.

## SUPPORTING INFORMATION

### Supporting Material and Methods

Word file

### Supporting Data S1:

Soil evaporation model.

### Figure S2:

Sampling and leaf water potential measurements timing. A, Greenhouse experiment: leaf harvest for transcriptomic occurs at day1 before artificial night and pre-dawn leaf water measurement. B, Field trials: Leaf water potential was measured before leaf harvest for transcriptomic data. C, Diurnal variation study: Pre-dawn leaf water potential was measured at day 1 at 4:00. Eight leaf water potential measurements were performed during the next 24 hours. At the same time of each measurement of $\Psi$ or $\Psi$PD leaves were harvested for transcriptomic study.

### Table S3

List of selected genes and their functional annotations.

Genes from Rengel *et al.* 2012, circadian clock genes, sunflower dehydrins and reference genes (Excel file).

### Table S4

Primers (Excel file).

**Figure S5**

Comparison ln(-ΨPD ) with biomarker prediction of ln(- ΨPD ) for the four best models. Red points show comparison between model prediction and raw observation of ΨPD observed in field experiments. Blue points show comparison between model prediction and corrected observations similar to ΨPD' used for model calibration.

**Figure S6**

Raw data of gene expression level function of FTSW values.The histogram shows average gene expression level for the corresponding FTSW level. Inedi is represented in black, PSC8 in blue, XRQ in red and Melody in green. The error bars represent the standard deviation for each genotype and each FTSW condition. The scatter plot shows gene expression level in function of FTSW value. One point represents one individual plant.

**Figure S7.**

P-value distribution for the correlation between expression level of the 28 candidate genes and the four WSI. The red line show the threshold selection p-value < 0.001

**Figure S8.**

P-value distribution of F-test for comparison between correlation model independent of genotypes and correlation model dependent of genotypes between candidate gene expression and the four WSI. The red line show threshold selection : p-value > 0.001.

**Table S9**

Soil analysis (Excel file).

**Figure S10**

Kinetic curves of circadian gene expression over 24 hours. Dotted lines show gene expression over 24 hours for 12 plots representing two genotypes (Inedi and Melody) and two conditions (irrigated and non-irrigated). A. Transcript HaT13l000567, homologous to the circadian Arabidopsis gene *ZTL*. B, Transcript HaT13l007116, homologous to the Arabidopsis circadian gene *TOC1*. C, Transcript HaT13l011336, homologous to the Arabidopsis circadian gene *LHY*. D, Transcript HaT13l015763, homologous to the Arabidopsis circadian gene *CO*. E,  Transcript HaT13l005099, homologous to the sunflower dehydrin HaDN1 (Cellier et al., 2000). F, Transcript HaT13l011509, homologous to the sunflower dehydrin HaDN2 (Cellier et al., 2000).

**Table S11**

ANOVA results for the trial effect. (Excel file)